\documentclass{vgtc}                          

\ifpdf
  \pdfoutput=1\relax                   
  \usepackage{graphicx}                

\usepackage{lipsum}

\RequirePackage[absolute]{textpos}
\DeclareRobustCommand*{\copyrightnotefirst}[1]{%
  \begin{textblock}{10}(28,243)
       \text{#1}%
  \end{textblock}%
    }

\DeclareRobustCommand*{\copyrightnotesecond}[1]{%
  \begin{textblock}{10}(20,247)
       \text{#1}%
  \end{textblock}%
    }

\DeclareRobustCommand*{\copyrightnotethird}[1]{%
  \begin{textblock}{10}(20,251)
       \text{#1}%
  \end{textblock}%
    }
  \DeclareGraphicsExtensions{.pdf,.png,.jpg,.jpeg} 
\else
  \usepackage{graphicx}                
  \DeclareGraphicsExtensions{.eps}     
\fi%

\graphicspath{{figures/}{pictures/}{images/}{./}} 

\usepackage{microtype}                 
\PassOptionsToPackage{warn}{textcomp}  
\usepackage{textcomp}                  
\usepackage{mathptmx}                  
\usepackage{times}                     
\usepackage{cite}                      
\usepackage{tabu}                      
\usepackage{booktabs}                  
\usepackage{amsmath}

\onlineid{0}

\vgtccategory{Research}

\vgtcinsertpkg

\title{Identifying Challenges in Designing, Developing and Evaluating \\ Data Visualizations for Large Displays}

\author{Mahsa Sinaei Hamed\thanks{e-mail: mahsasinaeihamed@cmail.carleton.ca}\\ %
        \scriptsize Carleton University %
\and Pak Kwan\\ %
     \scriptsize University of Cincinnati %
\and Matthew Klich\\ %
     {\scriptsize University of Cincinnati}
\and Jillian Aurisano\\ %
     {\scriptsize University of Cincinnati}
\and Fateme Rajabiyazdi\\ %
     {\scriptsize Carleton University}}

\abstract{
With the growth of data sizes, visualizing them becomes more complex. Desktop displays are insufficient for presenting and collaborating on complex data visualizations. Large displays could provide the necessary space to demo or present complex data visualizations. However, designing and developing visualizations for such displays pose distinct challenges. Identifying these challenges is essential for researchers, designers, and developers in the field of data visualization. In this study, we aim to gain insights into the challenges encountered by designers and developers when creating data visualizations for large displays. We conducted a series of semi-structured interviews with experts who had experience in large displays and, through affinity diagramming, categorized the challenges.
} 

\begin{document}



\firstsection{Introduction} 

\maketitle
\copyrightnotefirst{Poster presented at IEEE Visualization Conference 2023 }
\copyrightnotesecond{22-27 October - Melbourne, VIC, Australia}
\copyrightnotethird{Copyright held by authors}

In today's data-driven world, the size and complexity of data continue to grow. Analyzing and making sense of this vast amount of data is challenging. 
Data visualization is a technique which empowers users to visually explore, analyze, and interpret vast data.

Visualizations representing large data sets can be complex and require multiple views to capture various aspects of the data. Desktop-sized displays are often small and may not have enough screen space to effectively present these complex data visualizations or support collaborative work. These limitations emphasize the need for specialized infrastructure, such as large displays (LDs),  to effectively present complex data visualizations. 
LDs create a continuous visual area, at least the size of the human body, with the view space extending beyond human sight and possibly beyond human reach~\cite{belkacem2022interactive}.

LD environments support visual data exploration by providing abundant space for displaying and juxtaposing multiple views produced during the exploratory process~\cite{aurisano2020many} and enable users to leverage movement and embodied cognition~\cite{jakobsen2015moving} for improved memory in data intensive tasks, and facilitating collaboration~\cite{Rajabiyazdi2015}.    
Despite LDs offering an excellent opportunity for complex data visualizations, designing and developing for them requires extra considerations and poses several challenges. These design considerations include determining how visualization design principles for LDs may need to differ from those used for desktop displays, understanding the relative effectiveness of different encodings on LDs, and identifying the type of data visualizations which benefit most from LD sizes~\cite{andrews2011information}.

Previous literature has also highlighted challenges in designing data visualizations for LDs, such as the lack of visualization libraries which address LD technical requirements, out-of-reach interface elements due to display size, users need to move in front of LD when using the visualizations ~\cite{belkacem2022interactive},
and users perceive visual elements on LDs differently due to varying viewing angles~\cite{bezerianos2012perception}. Furthermore, designing interactions for LD environments presents challenges. Recent research explores modalities beyond mouse and keyboard inputs, such as touch~\cite{agarwal2019viswall} and proxemic interaction~\cite{ballendat2010proxemic}.
 
However, there are no formal studies that have investigated challenges in designing, developing, and evaluating data visualizations for LDs. Thus, we filled this gap through semi-structured interviews with experts with experience in creating data visualizations for LDs to identify the challenges faced by them.
Our findings identified preliminary challenges for designers and developers creating data visualizations for LDs, including limited LD infrastructure resources, scaling designs to LDs, limited development resources, and evaluating visualizations with people. Understanding these challenges will allow us to develop guidance for resources and tools to enable design, development and evaluation with LDs.

\section{Methodology} 

\textbf{Study Design:} We conducted 1-hour semi-structured interviews with open-ended questions. The interviews were designed to explore participants perspectives on the challenges encountered in designing, developing, and evaluating data visualizations for LDs. We conducted the interview sessions online via the Microsoft Teams platform and recorded them for transcription purposes.

We targeted computer science and computer engineering experts with experience working on LDs and creating data visualizations for LDs.
We used purposeful sampling to identify potential participants and reached out to them through email.
We recruited 4 participants (1 Female, 3 Male) aged between 37 and 43. All participants were experts who had experience in designing and developing data visualizations for LDs, with 6 to 11 years of experience.

\textbf{Data Analysis:} All the authors independently analyzed the interview transcripts using the affinity diagram technique~\cite{hartson2012ux}. We each identified the list of challenges mentioned by participants in the interviews on sticky notes on a Miro board. Then we collaboratively grouped sticky notes based on their similarities and relationships through group discussions. This process allowed us to group these challenges and identify patterns that emerged from the data. 

\section{Preliminary Findings}
The findings from our interviews shed light on the challenges encountered by experts when designing and developing data visualizations for LDs. We categorized the challenges into the following groups: 

\subsection{Limited Resources for Building LD Infrastructure}

Building LD infrastructure requires a large space to enable users to freely move and interact. This need imposed constraints in building and working with LDs. One participant (P1) mentioned \textit{``We moved all the chairs and tables because my research was focusing on making [minimum] three people to collaborate and use the large display so they can explore the data.''} 
In addition to space allocation, we need to implement effective cooling systems to manage the heat generated during LD operations. Prolonged use of LDs, especially in environments with inadequate ventilation or cooling, can lead to increased heat buildup, which could result in an uncomfortable working environment.  
To address these issues, we need to allocate sufficient space and install proper ventilation systems, which can be time-consuming and costly. As a result, this infrastructure typically involves fixed configurations that cannot be easily relocated.

Sharing the data visualization systems designed for LDs with the scientific community and the public is another challenge mentioned by our participants. 
Since moving LDs for demoing purposes is not easily possible and data visualizations designed for LDs cannot be easily demonstrated on a laptop, these demos are often shared in video format, which limits their accessibility and presentability.

The inability to easily relocate LDs and their associated infrastructure presents additional challenges when changing institutions. Our participants mentioned they lost access to LDs and technical and infrastructure-specific knowledge they had developed over the years when they changed their institutions. 

\subsection{Scaling Data Visualization Designs for LDs}

To design a data visualization prototype, we often start by sketching the designs. LDs often have different sizes, resolutions, and aspect ratios compared to paper or surfaces typically used for sketching. This disparity made it challenging for participants to sketch their data visualizations for LDs. Participants highlighted the need for flip-charts to create a fake wall, catering to the specific characteristics of LDs, in order to create sketches of their design on it. P3 said \textit{``what we did, which would just get flipchart paper, kind of corresponded to what we had in the screen wise for one screen. So we could take the whole wall with it and we had the wall that was big enough to actually simulate [the LD] \ldots
So later, we had a wall and like 4 flipchart stands as well so that you can do the curve thing as well. It was actually very effective.''}

\subsection{Limited Development Resources and Tools for LDs}

Our participants often developed prototypes away from LDs on their personal computers. This setup requires constant back-and-forth checks between personal computers and LDs to ensure correct size, spacing, and aspect ratio alignment, which was a time consuming process for developing data visualizations.

Unlike desktop-sized displays, LDs often require specialized interaction techniques to manipulate and explore the visualized data. Participants mentioned designing and implementing interactions for LDs was challenging for them as the interactions developed on their laptops were not compatible with the large screen size.

Moreover, participants mentioned they often needed to transfer data between different devices, such as smartphones or tablets and LDs, during the prototyping process. Coordinating these interactions and synchronizing the display across multiple devices required additional efforts, pointed out by our participants.

In addition to all of these challenges, available tools, software, and toolkits specifically tailored for LD prototyping is limited. While there may be some existing solutions, their functionalities and capabilities may not fully meet the requirements and challenges posed by LDs. Due to the limited availability of support and resources, developers often need to adapt or modify existing tools or create custom solutions to meet the requirements of LD prototyping.
P4 said \textit{``I didn't get too much help from  \ldots just in that because this is like a very specific platform that I needed.''}

\subsection{Evaluating Data Visualizations on LDs with People}

Evaluating data visualizations on LDs with people can be challenging as people often are unfamiliar with LD technology, as it is not widely accessible or commonly used. 
P3 said \textit{``We grew up with this thing that you don't want to break the TV. It's expensive and people are afraid to [touch] them. I think this is a challenge now [and it will not be a challenge] 20 years in the future.''} 

Our participants reported that their study users struggled with switching between the keyboard and the screen during the evaluation process, which disrupted the study workflow and impacted their productivity. In another case, a participant told us that their study users frequently experienced difficulties in keeping track of the cursor during the evaluation process.

Occasionally, we create a LD by combining desktop-sized displays. Navigating multi-display setups posed additional challenges during the evaluation process. Participants highlighted that navigating across multiple displays using a mouse required frequent hand movements, which was cumbersome for their study participants. 

Unlike controlled lab settings, public settings introduce various unpredictable challenges that can impact the evaluation process. One of our participants conducted evaluation studies in the wild to investigate effective techniques for capturing people's attention toward their work. They often faced additional challenges due to the lack of control over the experiment environment, which made it challenging to gather accurate data and observations.

\section{Discussion}
Our findings highlighted a preliminary list of challenges faced by designers and developers when creating data visualizations for LDs. These challenges include limited resources for building LD infrastructure, difficulties in scaling the designs to LDs, limited development resources, and challenges in evaluating data visualizations with people. Understanding these challenges could help us develop guidance for LD design, development, and evaluation resources and tools. This study underscores the importance of adopting tailored approaches and design processes to address these challenges.


\acknowledgments{
This work was funded by the Canadian NSERC fund RGPIN-2021-04222. We would like
to thank our participants for the expert knowledge they brought to this project.}

\bibliographystyle{abbrv-doi}

\bibliography{main}

\begin{thebibliography}{1}

\bibitem{agarwal2019viswall}
M.~Agarwal, A.~Srinivasan, and J.~Stasko.
\newblock Viswall: Visual data exploration using direct combination on large
  touch displays.
\newblock In {\em 2019 IEEE Visualization Conference (VIS)}, pp. 26--30. IEEE,
  2019.

\bibitem{andrews2011information}
C.~Andrews, A.~Endert, B.~Yost, and C.~North.
\newblock Information visualization on large, high-resolution displays: Issues,
  challenges, and opportunities.
\newblock {\em Information Visualization}, 10(4):341--355, 2011.

\bibitem{aurisano2020many}
J.~Aurisano, A.~Kumar, A.~Alsaiari, B.~D. Eugenio, and A.~Johnson.
\newblock Many at once: Capturing intentions to create and use many views at
  once in large display environments.
\newblock In {\em Computer Graphics Forum}, vol.~39, pp. 229--240. Wiley Online
  Library, 2020.

\bibitem{ballendat2010proxemic}
T.~Ballendat, N.~Marquardt, and S.~Greenberg.
\newblock Proxemic interaction: designing for a proximity and orientation-aware
  environment.
\newblock In {\em ACM International Conference on Interactive Tabletops and
  Surfaces}, pp. 121--130. ACM, 2010.

\bibitem{belkacem2022interactive}
I.~Belkacem, C.~Tominski, N.~M{\'e}doc, S.~Knudsen, R.~Dachselt, and
  M.~Ghoniem.
\newblock Interactive visualization on large high-resolution displays: A
  survey.
\newblock {\em arXiv preprint arXiv:2212.04346}, 2022.

\bibitem{bezerianos2012perception}
A.~Bezerianos and P.~Isenberg.
\newblock Perception of visual variables on tiled wall-sized displays for
  information visualization applications.
\newblock {\em IEEE Transactions on Visualization and Computer Graphics},
  18(12):2516--2525, 2012.

\bibitem{hartson2012ux}
R.~Hartson and P.~S. Pyla.
\newblock {\em The UX Book: Process and guidelines for ensuring a quality user
  experience}.
\newblock Elsevier, 2012.

\bibitem{jakobsen2015moving}
M.~R. Jakobsen and K.~Hornb{\ae}k.
\newblock Is moving improving?: Some effects of locomotion in wall-display
  interaction.
\newblock In {\em Proceedings of the 33rd Annual ACM Conference on Human
  Factors in Computing Systems}, pp. 4169--4178. ACM, 2015.

\bibitem{Rajabiyazdi2015}
F.~Rajabiyazdi, J.~Walny, C.~Mah, J.~Brosz, and S.~Carpendale.
\newblock Understanding researchers' use of a large, high-resolution display
  across disciplines.
\newblock In {\em Proceedings of the 2015 International Conference on
  Interactive Tabletops \& Surfaces}, ITS '15, p. 107–116. ACM, 2015.

\end{thebibliography}
\end{document}